\documentstyle[prd,aps,epsfig,floats]{revtex}

\newcommand{\dfrac}{\displaystyle \frac}
\newcommand{\btodpn}{B^0\rightarrow D^{*-}p\bar{n}}

\begin{document}
\draft
\twocolumn[\hsize\textwidth\columnwidth\hsize\csname
@twocolumnfalse\endcsname

\title{Understanding $B\rightarrow D^{*-}N\bar{N}$ and Its
Implications}

\author{Chun-Khiang Chua, Wei-Shu Hou and Shang-Yuu Tsai\\}
\address{Department of Physics, National Taiwan University,
Taipei, Taiwan 10764, R.O.C.}
\date{\today}
\maketitle

\begin{abstract}

The CLEO Collaboration recently reported the observation of the
$B^0\rightarrow D^{*-}p\bar{n}$ mode at a rate only a factor of
4--5 lower than $B^0\rightarrow D^{*-}\rho^+$ and $D^{*-}\pi^+$.
We try to understand this with a factorization approach of
current produced nucleon pairs. 
The baryon weak vector current form factors are related 
by isospin rotation to nucleon electromagnetic form factors. 
By using $G^{p,n}_M$ measured from $e^+e^-\rightarrow \bar{N}N$ 
and $p\bar{p}\rightarrow e^+e^-$ processes,
assuming factorization of the $B^0\rightarrow D^{*-}$ transition
and the $p\bar n$ pair production, we are able to 
account for up to $60\%$ of the observed rate.
The remainder is argued to arise from the axial current.
The model is then applied to $B$ decays to other mesons plus
$p\bar{n}$ modes and $D^{*-}$ plus strange baryon pairs.

\end{abstract}

]
%
%
%
%


\section{Introduction}\label{Intro}

Following a suggestion by Dunietz~\cite{Dunietz:1998uz} that 
$B \to D N\bar{N}X$ decays
could be sizable, 
the CLEO Collaboration has recently reported
the first observation of such modes~\cite{Anderson:2001tz}:
\begin{eqnarray}
{\rm Br}(B^0\rightarrow D^{*-}p\bar{n})&=&
(14.5^{+3.4}_{-3.0}\pm2.7)\times 10^{-4},
\label{B2Dpn}\\
{\rm Br}(B^0\rightarrow D^{*-}p\bar{p}\pi^+)&=&(6.5^{+1.3}_{-1.2}
\pm1.0)\times 10^{-4}.
\end{eqnarray}
Although the decay final states are three or four-body, 
they are only a few times below corresponding two-body mesonic
modes~\cite{Groom:2000in} such as 
\begin{eqnarray}
{\rm Br}(B^0\rightarrow D^{*-}\rho^+)&=&(6.8\pm3.4)\times
10^{-3},
\label{B2Drho}\\
{\rm Br}(B^0\rightarrow D^{*-}\pi^+)&=&(2.76\pm0.21)\times
10^{-3}\,\,.
\end{eqnarray}
Since $D^{*-}$ creation carries away much energy,
the observed large rate of $B^0\rightarrow D^{*-}p\bar{n}$ 
supports the suggestion that enhanced baryon production is 
favored by reduced energy release on the baryon side \cite{Hou:2001bz}.
Thus, given the large rate of $B\to\eta^\prime+X_s$ decay 
where $p_{\eta^\prime} > 2$ GeV~\cite{Browder:1998yb}, the 
$B\to\eta^\prime\Lambda\bar p$ decay may be sizable~\cite{Hou:2001bz}
compared to charmless two body baryonic modes. 
Similar argument holds for $B\to\gamma\Lambda\bar p$ 
as implied by $B\to \gamma +X_s$. 
Since $\Lambda\to p\pi$ decay automatically provides spin information,
the observation of such charmless three (or more) body baryonic modes
involving $\Lambda$ baryons allows for a search program for
triple-product type of $CP$ and $T$ violating effects.
With this in mind, 
a better understanding of the $B^0\rightarrow D^{*-}p\bar{n}$ decay 
is not only worthwhile in its own right, 
it can also serve as an important
first step towards a more ambitious project
on charmless baryonic modes.

\begin{figure}[b!]
\centerline{\hskip0.3cm {\epsfxsize1.75in \epsffile{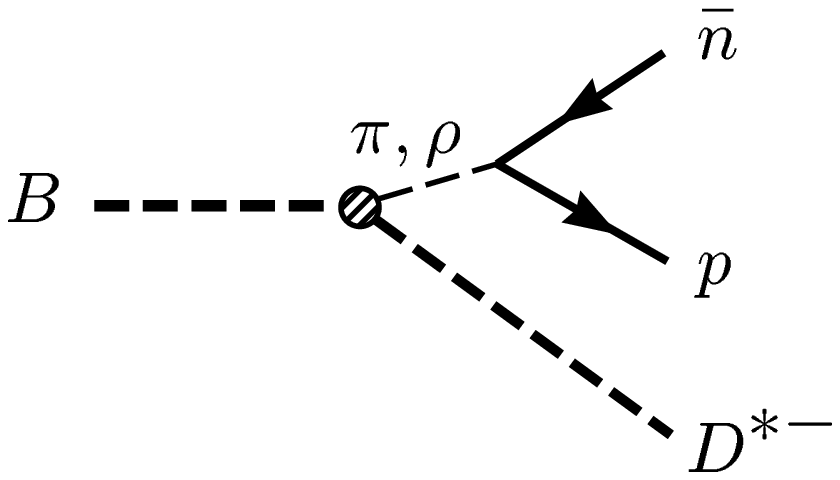}}
	   \hskip-0.5cm {\epsfxsize1.75in \epsffile{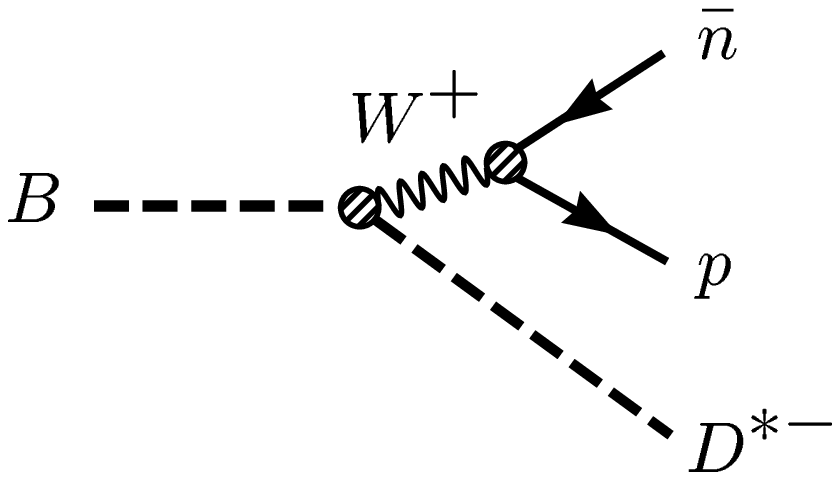}}}
 \smallskip  
\caption{
(a) The simple pole model and
(b) the factorized current model
for describing $B^0\rightarrow D^{*-}p\bar{n}.$
}
\label{Ansatz}
\end{figure}

In order to account for the proximity of rates shown in 
Eqs. (\ref{B2Dpn}) and (\ref{B2Drho}), 
$B^0\rightarrow D^{*-}p\bar{n}$ was seen through
a simple pole model~\cite{Hou:2001bz} as occurring in two steps
of an underlying $\bar b \to \bar c u \bar d$ transition,
i.e. $B\rightarrow D^{*-}h$ followed by $h\rightarrow N\bar{N}$, 
where $h$ stands for the off-shell $\pi^{+}$ and $\rho^{+}$ mesons,
as illustrated in Fig.~\ref{Ansatz}(a).
Taking this as a Feynman diagram, the decay amplitude of 
$B^0\rightarrow D^{*-}p\bar{n}$ can be written as 
\begin{eqnarray}
i{\cal{M}}_\rho 
&=& \left(-i\,\frac{G_F}{\sqrt{2}} V_{ud}V_{cb}^*\,a_1\right)
\langle D^{*-}|V_{\mu}-A_{\mu}|B^0\rangle
\nonumber\\
&& \quad\quad\quad\ \
\times \,f_{\rho}m_{\rho} \times i\biggl(\frac{-g^{\mu\lambda}
              +\frac{q^{\mu}q^{\lambda}}{m_{\rho}^2}} {q^2-m_{\rho}^2}
        \biggr) \nonumber\\
& 
\times & i\sqrt{2}\,
\bar{u}\bigl(p_p\bigr)\Biggl[g^{\rho N\bar N}_1
\gamma_{\lambda}
 +\frac{i\,g^{\rho N\bar N}_2}{2m_N}\sigma_{\lambda\xi}\,q^{\xi}\Biggr]
 \,v\bigl(p_{\bar{n}}\bigr),
 \label{amp}
\end{eqnarray}
where the $B^0\rightarrow D^{*-}$ transition matrix element is
\begin{eqnarray}
&&
\langle D^{*-}|V_{\mu}-A_{\mu}|B^0\rangle=
\Biggl[-\epsilon_{\mu\nu\alpha\beta}p^{\alpha}_B
p^{\beta}_{D^*}\frac{2V(q^2)}{m_B+m_{D^*}}
\nonumber \\
&& \quad\ \
-i\left(g_{\mu\nu}-\frac{q_{\mu}q_{\nu}}{q^2}\right)
(m_B+m_{D^*})A_1(q^2)
\nonumber\\
&& \quad\ \
 + i\left(\left(p_B+p_{D^*}\right)_{\mu}
  -\frac{m^2_B-m^2_{D^*}}{q^2}q_{\mu}\right)
   \frac{A_2(q^2)}{m_B+m_{D^*}}q_{\nu}
\nonumber\\
&& \quad\ \
-i\,2m_{D^*}\frac{q_{\mu}q_{\nu}}{q^2}
A_0\bigl(q^2\bigr)\Biggr] \, \epsilon^{*\nu}_{D^*},
\label{B2D}
\end{eqnarray}
$q=p_B-p_{D^*}=p_p+p_{\bar{n}}$ is the momentum transfer
(so $t = q^2$ is nothing but the $p\bar{n}$ pair mass),
$\epsilon_{D^*}$ is the polarization of the $D^{*-}$ meson, and
$g^{\rho N\bar N}_1$ and $g^{\rho N\bar N}_2$ are respectively
the vector (Dirac) and tensor (Pauli) coupling constants of the
$\rho$-meson to the nucleons.

In Eq.~(\ref{amp}),
factorization of the first vertex is justified to some extent,
and is expressed through the effective coefficient $a_1$.
For example, in ``naive factorization", $a_1=c_1+c_2/N_c$
where $N_c$ is the number of quark colors
and $c_{1,2}$ are Wilson coefficients,
but in general the effective coefficient $a_1$ can be extracted from 
experimental data on $B^0\rightarrow D^{*-}\rho^+$ decay.
The problem of the pole model picture is with 
single $\rho$ dominance and the $q^2$ dependence of the
strong interaction $\rho\,N\bar{N}$ vertex.
That is, 
there is no reason why the $u\bar d$ weak current only generates 
a $\rho$ meson, which then propagates and generates the baryon pair, 
as seen from the second and third lines of Eq.~(\ref{amp}).
Quantitative results based on this model is highly unreliable,
which is further aggravated by 
our ignorance of time-like meson-nucleon couplings. 
Since our knowledge of the coupling $g_1^{\rho NN}$ 
are based on low $Q^2(\equiv -q^2)$ space-like processes,
they cannot be expected to give reliable quantitative results
for time-like processes at higher energies.

In this paper we turn to a different approach by proposing 
a generalized factorization of current produced $p\bar{n}$ pairs. 
The three-body decay is seen as generated by 
two factorized weak currents (linked by a $W$-boson),
where one current converts $B^0$ into $D^{*-}$ and
the other creates the $p\bar{n}$ pair,
as shown in Fig.~\ref{Ansatz}(b).
In this way, having factorized the $B^0 \to D^{*-}$ transition,
we concentrate on the weak current production of baryon pairs.

It is well known that the vector portion of the weak current and 
the isovector component of the electromagnetic~(em) current form 
an isotriplet. 
Thus, information on the nucleon em form factors, for which 
much data exist in both the space-like region~\cite{eeScattering} 
(e.g. $ep\rightarrow ep$), and, 
of particular interest to us, the time-like 
region~\cite{Ambrogiani:1999bh,Antonelli:1998fv,Armstrong93,timelikeData}
(e.g. $e^+e^-\rightarrow N\bar{N}$ and $p\bar{p}\rightarrow e^+e^-$),
can be transferred to the nucleon weak form factors 
by a simple isospin rotation.
The total decay amplitude that comes from the vector portion of
the weak current can be written down unambiguously and the
portion of the branching fraction that comes solely from
the weak vector current can be readily
obtained once the form factors are given. 
We find that the vector current can account for 
50\%--60\% of the observed rate of Eq. (\ref{B2Dpn}).
Since this analysis involves just the factorization hypothesis
but is otherwise based on data, it is rather robust.

Although the axial vector and vector current contributions
interfere in the decay amplitude, the interference vanishes when one 
sums over spin and integrates over phase space.
The total rate is therefore a simple sum of the contributions from 
the vector and axial vector portions of the weak nucleon form factor. 
Like the vector case, 
we could in principle obtain the axial vector contribution 
if the nucleon form factors of the axial current were known.
Unfortunately, the time-like data is still lacking, hence the
contribution from this part remains undetermined. 
In spite of this, a rough estimate can still be given,
which seems to be the right amount. 
We point out, however, that information on 
the time-like {\it nucleon} axial form factor could 
in fact be obtained in the future via 
the $B^0\to D^{*-}p\bar{n}$ decay data.
One just has to separate the axial vector contribution 
from the vector part.

This paper is organized as follows: 
in the next section we
lay out the factorization assumption
that allows us to relate the vector current contribution to 
the nucleon em form factors,
where one enjoys abundance of experimental data.
We are then able to compute 
the vector current contribution to $B^0\to D^{*-}p\bar n$.
The axial vector contribution is also estimated. 
In Sec.~\ref{VMDapproach}, 
to illustrate the power of our data based approach,
we briefly introduce an improved pole model approach
and stress the need for improved measurements of neutron em form factors.
Finally, we apply our analysis to 
$B^+\to \bar{D}^{(*)0}p\bar{n}$ and $B^0\to D^-p\bar{n}$ and 
other baryonic modes containing strangeness,
and conclude in the last section.

\section{Factorization Approach and Nucleon Form Factor Data}
\label{fact}


As illustrated in Fig.~\ref{Ansatz}(b),
we generalize factorization from two-body
to three-body decay processes, 
\begin{eqnarray}
\langle D^{*-}p\bar{n}|{\cal H}_{eff}|B^0\rangle
&=&\frac{G_F}{\sqrt{2}}\,V_{ud}V_{cb}^*\,a_1\,\langle
D^{*-}|V^{\mu}-A^{\mu}|B^0\rangle \nonumber\\
&&\times \langle
p\bar{n}|V_{\mu}-A_{\mu}|\,0\rangle\,. \label{factorization}
\end{eqnarray}
The first matrix element is as before, 
but for the second, the nucleon pair is viewed
as directly created by the current.
The vector part can be expressed as
\begin{equation}
\langle p\bar{n}|V_{\mu}|0\rangle=
\bar{u}(p_p) \left\{F_1^W(t)\gamma_{\mu}+
   i\frac{\,F_2^W(t)}{2m_N}\sigma_{\mu\nu}q^{\nu}\right\}v(p_{\bar{n}}),
\label{weak_current}
\end{equation}
where $m_N$ is the nucleon mass, $t\equiv (p_p+p_{\bar n})^2 = q^2$,
and $F_{1,2}^W$ are the weak nucleon form factors. 
Likewise, the weak axial current $A_{\mu}\equiv A_{\mu}^1+iA_{\mu}^2$ 
matrix element is
\begin{eqnarray}
\langle p\bar{n}|A_{\mu}|0\rangle&=&
\bar{u}\left(p_p\right)\left\{g_A\left(t\right)\gamma_{\mu}
+\frac{h_A\left(t\right)}{2\,m_N}\,q_{\mu}\right\}\gamma_5\,
v\left(p_{\bar{n}}\right),\nonumber\\
&& \label{pAn}
\end{eqnarray}
where $g_A(t)$ is the axial form factor, and $h_A(t)$ is
referred to as the induced pseudoscalar form factor. 
Expressions for the space-like processes are similar.
The weak nucleon form factor $F^W_1(t)$ and the axial form
factor $g_A(t)$ are normalized at 
$t=0$~\cite{Groom:2000in}
\begin{equation}
F^W_1(0)=1,\qquad g_A(0)\equiv g_A=1.2670\pm0.0035\,.\label{norm1}
\end{equation}

\subsection{Isospin Relation and Nucleon Form Factors}\label{isospin}

It is well known that the photon field $A_\mu$ contains $W^3_\mu$,
which forms a weak isotriplet with $W^{1,2}_\mu$.
Therefore the currents they couple to also form an isotriplet, 
and are related by an isospin transformation. 
For the nucleon system, the strong isospin symmetry 
coincides with the weak isospin symmetry of the weak and em currents. 
The weak vector form factors are therefore related to 
isovector electromagnetic (em) form factors.

The matrix element $\langle N(p')\bar{N}(p)|{\cal J}^{em}_{\mu}|0\rangle$ 
for the em current can be expressed as
\begin{eqnarray}
&&\langle N(p')\bar{N}(p)|{\cal J}^{em}_{\mu}|0\rangle
\nonumber\\
&& 
=\bar{u}(p') \biggl\lbrace
F_1(t) \gamma_\mu + i\frac{F_2(t)}{2m_N}  \sigma_{\mu \nu}  q^\nu
\biggr\rbrace v(p). 
\label{em-current}
\end{eqnarray}
Similar expressions can be obtained for space-like processes.
The quantities $F_1 (t)$, $F_2(t)$ are respectively the Dirac and
Pauli form factors, normalized at $t=0$ as
\begin{equation}
F_1^p(0) = 1, \quad F_1^n(0) = 0, \quad 
F_2^p(0) = \kappa_p, \quad F_2^n(0) = \kappa_n,
 \label{norm2}
\end{equation}
with $\kappa_{p\ (n)}$ the proton (neutron) 
anomalous magnetic moment in nuclear magneton units. 
The experimental data is usually given in terms of 
the Sachs form factors, which are related to $F_1$ and $F_2$ through
\begin{eqnarray}
G^{p,n}_E(t)&=&F_1^{p,n}(t)+\frac{t}{4m_N^2}F_2^{p,n}(t)\,,\nonumber\\
G^{p,n}_M(t)&=&F_1^{p,n}(t)+F_2^{p,n}(t)\,.
\label{EMff}
\end{eqnarray}
One clearly has $G_M=G_E$ at threshold $t=4m_N^2$, 
while at $t=0$ we have
\begin{equation}
G^p_E(0)=1,\,\, G^n_E(0)=0,\,\,
G^p_M(0)=1+\kappa_p,\,\, G^n_M(0)=\kappa_n.
\label{norm}
\end{equation}

The isospin decomposition of the em current is given by the
following definitions
\begin{equation} F_i^{s,v} = \dfrac{1}{2}\, 
(F_i^p \pm F_i^n), \quad (i = 1,2),
 \label{isospin_decomp}
\end{equation}
where $F_i^{s}$ and $F_i^{v}$ are the isoscalar and isovector
decompositions of the form factors, respectively. The fact that the
isovector component of the em current, together with the vector
portion of the charged weak currents, form an isotriplet is
manifested by
\begin{equation}
2\,F_i^{v}(t)=F_i^W(t),\quad\quad\quad\quad\quad\quad
i=1,2, \label{WeakEMff}
\end{equation}
the factor $2$ coming from the definition of $F_{1,2}^{(s,v)}(t)$
in Eq.~(\ref{isospin_decomp}). 
For example, from Eqs.~(\ref{norm1}), (\ref{norm2}) 
and (\ref{isospin_decomp}) we have $2\,F_1^{v}(0)=F_1^W(0)$.

With Eq.~(\ref{WeakEMff}) and the Gordon decomposition, 
we can put the three-body $B^0\to D^{*-} p \bar n$ decay amplitude
of Eq.~(\ref{weak_current}) into the following form
\begin{eqnarray}
&&
i{\cal{M}}_V=  
-i\frac{G_F}{\sqrt{2}}
V_{ud}V^*_{cb}\,a_1\, \epsilon_{D^*}^{*\nu}
\Biggl[-\epsilon_{\mu\nu\alpha\beta}
p^{\alpha}_B p^{\beta}_{D^*}
 \frac{2V(q^2)}{m_B+m_{D^*}}
 \nonumber\\
&&
-ig_{\mu\nu}(m_B+m_{D^*})A_1(q^2)
+i\left(p_B+p_{D^*}\right)_{\mu} q_{\nu}
   \frac{A_2(q^2)}{m_B+m_{D^*}}\Biggr]
\nonumber\\
&&
\times \bar{u}(p_p) 
\Biggl[2\left(F^{v}_1+F^{v}_2\right) \gamma^{\mu}
 +\frac{F^{v}_2}{m_N} \left(p_{\bar{n}}-p_p\right)^{\mu}
\Biggr]v\left(p_{\bar{n}}\right),
 \label{amplitude}
\end{eqnarray}
where $V$ indicates that the nucleon pair is generated by 
the vector current.
%
%
The terms proportional to $q_\mu$ in Eq.~(\ref{B2D}) vanish in 
the limit of equal proton and neutron mass.
For completeness, the amplitude for nucleons generated by 
the axial vector current is given by
\begin{eqnarray}
&& i{\cal{M}}_A=   
-i\frac{G_F}{\sqrt{2}} V_{ud}V_{cb}^*\,a_1\,
\langle D^{*-}|V_{\mu}-A_{\mu}|B^0\rangle
\nonumber\\
&&  \quad\ \quad\
\times \bar{u}\left(p_p\right)
\Biggl[g_A\left(t\right)\gamma^{\mu}\gamma_5
+\frac{h_A\left(t\right)}{2\,m_N}\,q^{\mu}\gamma_5\Biggr]
v\left(p_{\bar{n}}\right),
 \label{axial}
\end{eqnarray}
where $\langle D^{*-}|V_{\mu}-A_{\mu}|B^0\rangle$ is
given in Eq.~(\ref{B2D}).

The two amplitudes ${\cal M}_A$ and ${\cal M}_V$ interfere, since
\begin{eqnarray}
&&
\sum 2{\rm Re}\left({\cal M}_A{\cal M}^*_V\right)=
32\,G_F^2\,|V_{ud}|^2\,|V_{cb}|^2\, a_1^2\,
\nonumber\\
&&  \quad\quad\quad\quad\quad\quad
\times g_A(t)V(t)A_1(t) \Bigl(G^p_M(t)-G^n_M(t)\Bigr)
\nonumber\\
&&  \quad\quad\quad
\times\Bigl[(p_B\cdot p_p)(p_{D^*}\cdot
p_{\bar{n}})-(p_{D^*}\cdot p_p)(p_B\cdot p_{\bar{n}})\Bigr]
\label{asymmetry}
\end{eqnarray}
is in general non-vanishing. 
The summation is performed over the nucleon spins
and $D^{*-}$ polarization.  
The three-body phase space is described by 
the two independent variables 
$m_{p\bar n}^2\equiv t=(p_p+p_{\bar{n}})^2$
and $m_{{D^*}\bar{n}}^2\equiv (p_{D^*}+p_{\bar{n}})^2$. 
The interference term is antisymmetric with respect to 
exchange of $p_p$ and $p_{\bar{n}}$.
For given $t$, as we integrate over the kinematically
allowed $m_{{D^*}\bar{n}}^2$, 
each value of $\sum 2Re({\cal M}_A{\cal M}_V^*)$ 
from a given pair of $p_p$, $p_{\bar{n}}$ would be
cancelled by those from the exchanged pair.
%
The interference term thus contributes nothing to the total three
body decay rate $\Gamma$, and the final result is a simple sum
of the contribution from ${\cal M}_V$ and ${\cal M}_A$ separately.

It is interesting to note that, although the effect of 
$\sum 2Re({\cal M}_A{\cal M}_V^*)$ does not show up in the decay rate,
we can nevertheless construct an asymmetry ratio measurable based on
the antisymmetric nature of this quantity, to extract the
time-like information of $g_A(t)$ from $B^0\to D^{*-} p \bar n$ data.
The information of $V(t)$, $A_1(t)$ and $G^{p,n}_M(t)$ in
Eq.~(\ref{asymmetry}) can be found by other means, 
and the overall factors $a_1$ and $|V_{ud}|$, $|V_{cb}|$ 
would cancel in the asymmetry ratio.

\subsection{Form Factor Data and Perturbative QCD}\label{pQCD}

Much data has been accumulated for the nucleon em form factors,
which turns the vector portion of the decay amplitude $i{\cal M}_V$
expressed in Eq.~(\ref{amplitude}) into something that we can handle.
The branching fraction can be readily obtained
once the nucleon em form factors are given, thanks to the isospin
relation of weak vector and em currents. 
It is important, however, to make sure that the form factors 
satisfy the constraint from perturbative QCD~(PQCD).

The so-called quark counting
rules~\cite{Brodsky:1975vy} give the leading power in the
large-$|t|$ fall-off of the form factor $F_1(t)$ by counting the
number of gluon exchanges which are necessary to distribute the
large photon momentum to all
constituents.
Since helicity-flip leads to an extra $1/t$ factor in $F_2(t)$,
one finds, in the limit $|t|\rightarrow \infty$
\begin{eqnarray}
F_i (t) \to (|t|)^{-(i+1)} \biggl[
\ln\biggl(\dfrac{|t|}{Q_0^2}\biggr) \biggr]^{-\gamma}, &&\quad
\gamma = 2 + \dfrac{4}{3\beta},\nonumber\\
&& \,\quad i = 1,2
 \label{asym}
\end{eqnarray}
where $\beta$ is the $\beta$--function of QCD to one loop,
and $Q_0 \simeq \Lambda_{\rm QCD}=0.3$ GeV. 
We note that $\gamma$ depends weakly on the number of flavors;
for three flavors $\gamma=2.148$.

The asymptotic form given in Eq.~(\ref{asym}) has been confirmed
by many experimental measurements of $G_M=F_1+F_2$ over a wide range
of momentum transfers in the space-like region.
The asymptotic behavior for $G^p_M$ also seems to hold in 
the time-like region, as reported by the Fermilab 
E760 experiment~\cite{Armstrong93} for $8.9$~GeV$^2<t<13$~GeV$^2$. 
Another Fermilab experiment, E835,
has recently reported $G^p_M$ for 
momentum transfer up to $\sim 14.4$~GeV$^2$,
and gives the empirical fit~\cite{Ambrogiani:1999bh}:
\begin{equation}
\left|G^p_M\right|=
\frac{C}{t^2\left[\,\ln\left(\frac{t}{Q_0^2}\right)\right]^2},
\label{empirical}
\end{equation}
which agrees well with the asymptotic form in Eq.~(\ref{asym}).

\begin{figure}[t!]
\begin{center}
\epsfig{figure=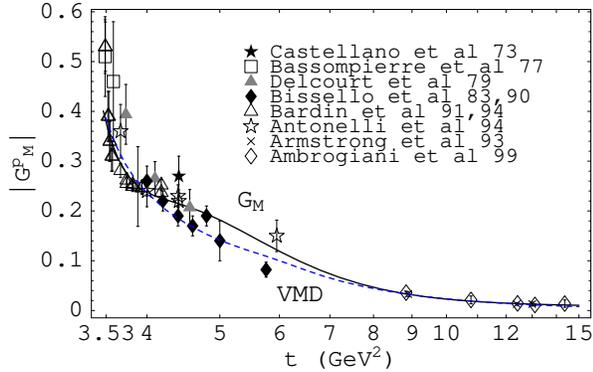,width=3.3in}\end{center}
\caption{Time-like proton magnetic form factor data, 
fitted (solid) by Eq.~(\ref{fitP}) 
with the parameters given in Eq.~(\ref{bestP}).
The other~(dash) curve is discussed in Sec. III.}
\label{proton}
\end{figure}

By exploiting the relation in Eq.~(\ref{EMff}), the combination
$2(F^v_1+F^v_2)$ in Eq.~(\ref{amplitude}) can be replaced by
$G^p_M-G^n_M$ which is composed of measurable quantities. 
Similar replacement can also be made for $F_2$, 
which is a combination of $G^p_M-G^p_E$ and $G^n_M-G^n_E$. 
Most time-like data for the magnetic form factors, however, 
are extracted by assuming either $|G_E|=|G_M|$ or $|G_E|=0$ 
in the region of momentum transfer explored. 
Since $G_M - G_E = (1 -{t}/{4m_N^2}) F_2$ clearly vanishes at threshold,
the absence of clear deviation from this assumption in extracting data
implies the contribution of $F_2$ is negligible 
even for $t$ far beyond the threshold, 
which is consistent with the prediction from QCD. 
In fact, by assuming $|G_E|=|G_M|$ in extracting $G_M$ from data, 
the information on $F_2$ is lost.
In our calculation we concentrate on the part of
Eq.~(\ref{amplitude}) which contains $F^v_1+F^v_2$.
The contribution from $F^v_2$ can be determined 
only when $G_M$ and $G_E$ can be separated from data 
with better angular resolution.

\begin{figure}[t!]
\begin{center}
\epsfig{figure=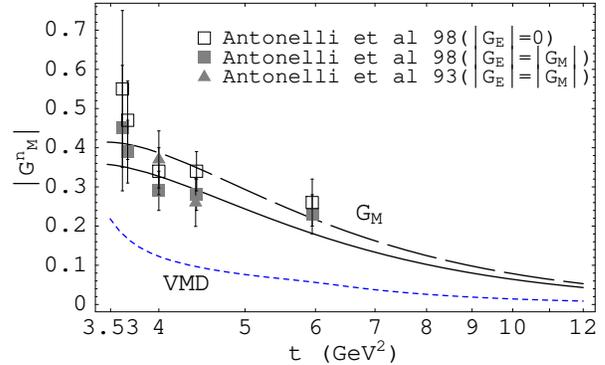,width=3.3in}
\end{center}
\caption{
Time-like neutron magnetic form factor, where
solid (long dash) line is fitted by Eq.~(\ref{fitN}) 
with parameters given in Eq.~(\ref{bestNe}) (Eq.~(\ref{bestN})), 
for data extracted with $|G_E|=|G_M|$ ($|G_E|=0$) assumption. 
The third~(dash) curve is discussed in Sec. III.
}
\label{neutron}
\end{figure}

We take $|G_M|$ in the following form to make 
a {\it phenomenological fit} of the experimental 
data~\cite{Ambrogiani:1999bh,Antonelli:1998fv,Armstrong93,timelikeData}:
\begin{equation}
\left|G^p_M(t)\right|=\left(\frac{x_1}{t^2}
+\frac{x_2}{t^3}+\frac{x_3}{t^4}
+\frac{x_4}{t^5}+\frac{x_5}{t^6}\right)\left[\ln\left(\frac{t}{Q_0^2}
\right)\right]^{-\gamma},
 \label{fitP}
\end{equation}
%
%
\begin{equation}
\left|G^n_M(t)\right|=\left(\frac{y_1}{t^2}+\frac{y_2}{t^3}\right)
\left[\ln\left(\frac{t}{Q_0^2}\right)\right]^{-\gamma},
 \label{fitN}
\end{equation}
where the leading power and log factor are from Eq.~(\ref{asym}), 
and the fewer parameters for $G_M^n$ reflects 
the scarcer amount of neutron data. 
We find the best fit values
\begin{eqnarray}
&x_1&=420.96~\mbox{GeV}^4,
\,\,\,\quad\quad\quad x_4=-433916.61~\mbox{GeV}^{10},\nonumber\\
&x_2&=-10485.50~\mbox{GeV}^6,
\quad\quad x_5=613780.15~\mbox{GeV}^{12}, \nonumber\\
&x_3&=106390.97~{\rm GeV}^8,
 \label{bestP}
\end{eqnarray}
and
\begin{equation}
y_1=236.69~{\rm GeV}^4,\quad \quad \quad y_2=-579.51~{\rm
GeV}^6,
 \label{bestNe}
\end{equation}
where the $\chi^2$ per degree of freedom of the fits are 
$1.47$, $0.41$ for $|G^p_M|$, $|G^n_M|$, respectively. 
We show in Figs.~\ref{proton} and \ref{neutron} the fitted data
together with the best fit curves given by 
Eq. (\ref{fitP}) and (\ref{fitN}) with the parameters above.

It was pointed out in Ref.~\cite{Antonelli:1998fv} that the
data supports $|G^n_E|=0$ as well. 
We therefore perform the fit
for the neutron magnetic form factor to the data that is
extracted under the assumption of $|G^n_E|=0$. 
Since the number of data points is small, 
a two-parameter fit as in Eq.~(\ref{fitN}) would still suffice. 
The best fit values are
\begin{equation}
y_1=292.62~{\rm GeV}^4\,,\quad \quad \quad y_2=-735.73~{\rm
GeV}^6\,, \label{bestN}
\end{equation}
giving $\chi^2/$d.o.f. $=0.39$, which is 
a little smaller than the previous fit,
and the fit is plotted as the long-dashed line in Fig.~\ref{neutron}. 
To the eye, the data might slightly prefer the second fit,
especially the $t = 6$ GeV$^2$ data point. 
However, it should be clear that more data is needed to
distinguish between these two cases.

We note that there is a sign difference between $G^p_M$ and
$G^n_M$ in the space-like region, since from Eq.~(\ref{norm})
$G^p_M(0)=1+\kappa_p>0$ and $G^n_M(0)=\kappa_n<0.$
Analyticity implies continuity at infinity 
between space-like and time-like $t$~\cite{Logunov} regions.
Hence time-like magnetic form factors are expected to behave like
space-like magnetic form factors, 
i.e. real and positive for the proton, but negative for the neutron. 

For large $t$, QCD predicts 
the magnetic form factors to be real~\cite{Brodsky:1975vy}, 
with the neutron form factor weaker than the proton case~\cite{Matveev73}.
According to QCD sum rules~\cite{Chernyak:1984bm}, 
asymptotically one expects $G^n_M/G^p_M\sim Q_d/Q_u=-0.5.$ 
We can readily check our fits against these: 
for $G_M$ fitted to data extracted by assuming $|G_E|=|G_M|$ 
for both neutron and proton magnetic form factors, we have
$G^n_M/G^p_M=-y_1/x_1=-236.69/420.96=-0.56\sim -0.5,$. 
For $G_M$ fitted to the proton data extracted with the assumption
$|G_E|=|G_M|$ but the neutron data extracted assuming $|G_E|=0$,
we have $G^n_M/G^p_M=-y_1/x_1=-292.62/420.96=-0.70,$ slightly
larger than $-0.5$.
Nucleon form factors have also been analyzed
from negative to positive $t$ with dispersion relations. 
The phase of the proton magnetic form factor turns out to be 
$\sim 2\pi$, hence the proton magnetic form factor is
real and positive as expected asymptotically, but already from
$t\ge 4$~GeV$^2$~\cite{Baldini:1999qn,Hammer:1996kx} onwards.

\subsection{Results for $B^0\to D^{*-} p\bar n$}

Before using data and the nucleon form factor relations 
to compute the decay rate,
we need to specify the value of $a_1$ to be used.
Since Eq. (\ref{amplitude}) depends on 
the product of $a_1$ and $B\to D^*$ form factors~\cite{Ali:1998eb},
we take a phenomenological approach
and use the value of $a_1$ extracted from 
the two-body mode $B\rightarrow D^{*-}\rho^+$,
i.e. 
$a^{{\rm BSW}}_1=0.86\pm0.21\pm0.07$ and
$a^{{\rm LF}}_1=0.74\pm0.18\pm0.06$ 
for Bauer-Stech-Wirbel (BSW)~\cite{Wirbel:1985ji,Bauer:1987bm}
and light-front (LF) form factor models~\cite{Cheng:1997if}.
The CKM matrix elements $V_{ud}$ and $V_{cb}$ are taken to
be $0.9757$, $0.039$ respectively.

For both proton and neutron data extracted by
assuming $|G_E|=|G_M|$, the predicted branching ratios are
\begin{equation}
\mbox{Br}_V^{{\rm LF}}=(7.14^{+0.69}_{-0.65})\times
10^{-4}\left(\frac{a_1}{0.74}\right)^2,
 \label{LFBr}
\end{equation}
for the light-front~(LF) model, and
\begin{equation}
\mbox{Br}_V^{{\rm BSW}}=(8.72^{+0.85}_{-0.79}) \times
10^{-4}\left(\frac{a_1}{0.86}\right)^2,
 \label{BSWBr}
\end{equation}
for the BSW model. The subscript $V$ serves as a reminder that
this is the result from the vector portion of the weak current
alone. The upper and lower errors correspond respectively to the
maximum and minimum of the branching fraction evaluated by
scanning through both $\chi^2\leq\chi^2_{\rm min}+1$ of $|G^p_M|$
and $|G^n_M|$ fits.

\begin{figure}[b!]
\begin{center}
\epsfig{figure=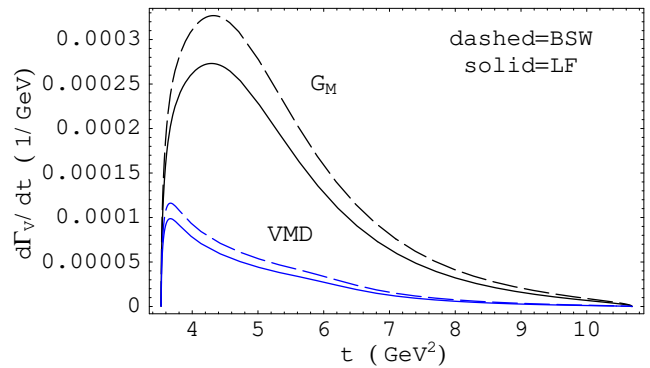,width=3.3in}
\end{center}
\caption{The differential decay rate vs $t = m_{p\bar n}^2$ 
of the vector current induced $B^0\rightarrow D^{*-}p\bar{n}$ decay.
The upper curves are from the phenomenological fits to 
nucleon form factor data assuming $|G_E|=|G_M|$, 
for the Light-Front~(solid) and BSW~(dashed) 
$B^0\rightarrow D^{*-}$ form factor models.
The lower curves are for the VMD model discussed in the next section.
} \label{dGdq}
\end{figure}

For proton data extracted by assuming $|G_E^p|=|G_M^p|$ but
neutron data assuming $|G_E^n|=0$, we have
\begin{eqnarray}
\mbox{Br}_V^{{\rm LF}}&=&(8.96^{+1.02}_{-0.94})\times
10^{-4}\left(\frac{a_1}{0.74}\right)^2,
 \label{LFBr0}
\\
\mbox{Br}_V^{{\rm BSW}}&=&(10.94^{+1.25}_{-1.15}) \times
10^{-4}\left(\frac{a_1}{0.86}\right)^2.
 \label{BSWBr0}
\end{eqnarray}
The larger values for this second set of branching fractions 
can be understood qualitatively from Fig.~\ref{neutron}, where 
the curve fitted to data assuming $|G_E|=0$ is higher than 
the one fitted assuming $|G_E|=|G_M|$.
Since the neutron magnetic form factor is 
negative in the time-like region, 
the quantity $G^p_M-G^n_M$ is larger 
if $G^n_M$ gets more negative,
and the branching fraction becomes larger.

Comparing with the central value of the measured 
Br$(\btodpn)=(14.5{+3.4\atop -3.0}\pm 2.7)\times 10^{-4}$,
our LF~(BSW) model results contribute $50\ (60)\%$ for
the first set and $62\ (75)\%$ for the second.
If the naive factorization vaue for $a_1$ is used, 
the results are close to the experimental central value, 
that is Br$_V^{{\rm LF(BSW)}}=12.51\ (13.83)\times 10^{-4}$ 
for the first set and $15.69\ (17.36)\times 10^{-4}$ for the second set. 
We plot in Fig.~\ref{dGdq} 
the vector current induced differential decay rate 
${d\Gamma_V\left(B^0\rightarrow D^{*-}p\bar{n}\right)}/{dq^2}$,
for both the LF and the BSW models.
The lower two curves are from the approach of next section.
As seen also from Eqs.~(\ref{LFBr}--\ref{BSWBr0}),
the LF form factor model gives results that are smaller
than the BSW model case.
The $\sim 10\%$ difference can be viewed as 
an estimate of the uncertainty from $B\to D^*$ form factor models.

From Fig.~\ref{dGdq} we see that the differential rate 
peaks at $\sim 4.6$ GeV$^2$, corresponding to
$m_{p\bar n} \simeq 2.14$ GeV, 
which is quite close to the threshold of 1.88 GeV. 
This threshold enhancement effect, 
consistent with what was suggested in Ref. \cite{Hou:2001bz},
should be checked experimentally 
by measuring the recoil $D^{*-}$ momentum spectrum.

\subsection{Estimate of Axial Current Contribution}

Although the time-like data for the form factors of the axial
current is still lacking, we can nevertheless make a rough
estimate of its contribution. 
In analogy with the nucleon em form factors which are 
constrained by the asymptotic form of Eq. (\ref{asym}), we expect that, 
for large $t$, $g_A\left(t\right)$ behaves as $1/t^2$ and 
dominates over $h_A\left(t\right)$, which behaves like $1/t^3$. 
Taking cue from the similarity between Eqs. (\ref{amplitude}) 
and (\ref{axial}), we estimate the axial vector contribution 
by making a simple comparison and scaling from the vector case. 
Since we only have space-like information for the $g_A(t)$ form factor, 
we estimate $g_A(t_{\rm th})$ at threshold by assuming same
threshold enhancement factor as the $G^p_M(t)-G^n_M(t)$ case.

With this and Eqs. (\ref{amplitude}) and (\ref{axial}) in mind,
to estimate the axial vector current contribution,
we scale the decay rate obtained from the vector case by the ratio 
$g^2_A(-t_{\rm th})/(G^p_M(-t_{\rm th})-G^n_M(-t_{\rm th}))^2$ 
for space-like momenum. 
We use a dipole fit $g_A(t)=g_A(0)/(1-t/M_A^2)^2$ 
with the axial mass $M_A = 1.077\pm0.039$ GeV~\cite{Liesenfeld:1999mv}. 
For $t=-4m_N^2$, the ratio 
$r(t)\equiv g_A\left(t\right)/(G^p(t)-G^n_M(t))$ 
gives $r(-4m_N^2)\sim 0.59$. 
Assuming this ratio holds also at the threshold $t=4m_N^2$, 
then $r^2(4m_N^2)\sim 0.35$ could be taken as 
the ratio of the branching fraction from the weak axial vector current 
to that from the weak vector current, 
i.e. ${\rm Br}_A/{\rm Br}_V\sim r^2(4m_N^2)$. 
For the results from fitting data assuming $|G_E|=|G_M|$, 
the total rate ${\rm Br}={\rm Br}_V+{\rm Br}_A$ would then be 
${\rm Br}^{{\rm LF}}\sim(1+0.35)\times (7.14\times 10^{-4})= 
9.64\times 10^{-4}$ and ${\rm Br}^{{\rm BSW}}\sim 11.77 \times 10^{-4}$. 
On the other hand, for the results from fitting data assuming $|G_E|=0$,
the total rate would be  ${\rm Br}^{{\rm LF}}\sim(1+0.35)\times
(8.96\times 10^{-4})= 12.10\times 10^{-4}$ and ${\rm Br}^{{\rm BSW}}\sim
14.77 \times 10^{-4}$, quite compatible with the experimental value of
$(14.5{+3.4\atop -3.0}\pm 2.7)\times 10^{-4}$.
Another value of $t$ which could be of interest is $t=0$. 
Assuming that the ratio $r(0)\sim 0.27$ holds at threshold, 
then from $r^2(0)\sim 0.07\sim {\rm Br}_A/{\rm Br}_V$, 
the total rate ${\rm Br}={\rm Br}_V+{\rm Br}_A$ is 
dominated by the vector current contribution. 

To improve our result, we urge for more experimental measurements of 
$G_M^n(t),\,G^{p,n}_E(t)$ and $g_A(t)$.
In turn, if the predictions (strength and spectrum) from our model 
based on the vector part are confirmed by experiment, 
the meaurements of $B\to \bar D^* p\bar n$
could provide useful feedback on nucleon axial form factors.

\section{Interrelation with Nucleon Form Factor Models}
\label{VMDapproach}

The nucleon form factor 
is one of the oldest subjects in particle physics.
It has provided us with a wealth of information and insight,
and remains an active field to this date.  
In the preceding section, we made a simple phenomenological fit 
of nucleon em form factor data, then used an isospin relation and 
a factorization hypothesis to compute the
$B^0\to D^{*-}p\bar n$ rate, with some success.
Although we made use of PQCD counting rules, 
we did not utilize tools such as analyticity.
On the other hand, we mentioned the possibility that
$B^0\to D^{*-}p\bar n$ type of modes could eventually 
provide useful information on the nucleon form factor itself.

To exploit the utility of analyticity
and to illustrate future interrelations between
$B^0\to D^{*-}p\bar n$ and nucleon form factors,
we adopt a specific nucleon form factor model
and discuss where it may be improved.
The discussion would also shed some light on form factor decompositions,
as well as the possible approach to 
$B\to \gamma p\bar\Lambda$, $\eta^\prime p\bar\Lambda$ modes,
which we shall briefly touch upon later.

\subsection{Dispersion Analysis and VMD Hypothesis}

\begin{figure}[t!]
\begin{center}
\epsfig{figure=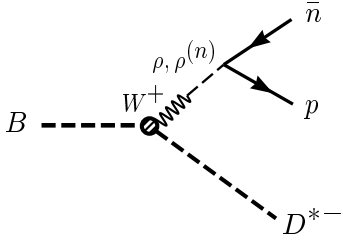,width=1.9in}
\end{center}
\caption{The VMD picture for $B^0\to D^{*-}p\bar n$ decay. 
The $W$ boson couples to the nucleon pair via vector mesons 
which in our case are $\rho$ and three excited states,
including a fictitious pole to mimic continuum effect.}
\label{Feyn_VMD}
\end{figure}

Among the various approaches to the nucleon em form factors, 
of particular interest is the Vector-Meson-Dominance (VMD) hypothesis,
which states that a photon couples to the hadrons via
intermediate vector mesons such as $\rho$, $\omega$, $\phi$, etc.
The simple pole model of Fig.~\ref{Ansatz}(a) 
is a limited form of the VMD hypothesis. 
In Ref.~\cite{Mergell:1996bf}, a parameterization of the nucleon em form
factors based on dispersion analysis was proposed, 
which is constrained by several physical conditions,
including PQCD power counting. 
The starting point is the dispersion relation
\begin{equation}
F(t) = \frac{1}{\pi} \, \int_{t_0}^\infty \frac{{\rm Im} \,
F(t')}{t'-t} \, dt', \label{disp}
\end{equation}
where $F(t)$ stands for the nucleon em form factors $F_{1,2}^{v}(t)$.
To gain predictive power, one truncates the spectral function
by a few vector meson poles,
where, to mimic the effect of large $t$ continuum
and to enforce PQCD counting rules,
a fictitious pole is introduced.
Thus, the form factors take the form
\begin{equation}
F^{v}_i(t)=\sum_v\frac{a^{v}_i}{M_{v}^2-t},
 \,\,\,\,\,\,\,\,\,\,\,\,\,\,i=1,2,
\label{VMD}
\end{equation}
where $a_{1,2}^{v}$ are related to the vector~($i=1$) and
tensor~($i=2$) coupling constants of Eq. (\ref{amp}) via
\begin{equation}
\sqrt{2}\,a^{v}_i=m_{v}f_v \,g^{vNN}_i,
\label{z}
\end{equation}
and $f_v$ is the vector meson decay constant.
This clearly extends the simple $\rho$ meson exchange picture. 
Together with our factorization ansatz,
the $B^0\to D^{*-}p\bar n$ transition can be 
pictured as in Fig.~\ref{Feyn_VMD}.

Following Ref.\cite{Mergell:1996bf} we now consider the following
parameterization of the form factors:
\begin{eqnarray}
&&F_i^{v} (t)
 = \biggl[ \tilde{F}_i^\rho (t) + \sum_v \dfrac{a_i^{v}
L^{-1}(M^2_{v})}{M^2_{v} - t }\biggr]
L(t),
\label{ffivm}
\end{eqnarray}
where
\begin{equation}
L(t) \equiv \biggl[ \ln \biggl(
\dfrac{\Lambda^2 - t}{Q_1^2} \biggr)\biggr]^{-\gamma},
\label{deflt}
\end{equation}
\begin{eqnarray}
\tilde{F}_i^\rho (t) 
&=& \dfrac{ a_i^\rho L^{-1}(M_a^2) + b_i^\rho L^{-1}(M_b^2) \bigl(
1 - t / c^\rho_i \bigr)^{-2/i}}{2 \bigl( 1 - t / d^\rho_i \bigr)},
\nonumber\\
&&
\label{lt}
\end{eqnarray}
for $i=1,2$, where
$a_1^{\rho}=1.0317$, $a_2^{\rho}=5.7824$, $b_1^{\rho}=0.0875$,
$b_2^{\rho}=0.3907$, $c_1^{\rho}=0.3176$, $c_2^{\rho}=0.1422$,
$d_1^{\rho}=0.5496$, $d_2^{\rho}=0.5362$, and $M_a^2=0.5$~GeV$^2$,
$M_b^2=0.4$~GeV$^2$.
Since the form factors also receive constraints from perturbative QCD
for large momentum transfer, a logarithmic factor,
Eq. (\ref{deflt}), is given for consistency.
Apart from this, Eq. (\ref{ffivm}) contains two terms: 
$\tilde{F}_i^\rho (t)$ represents 
the $2\pi$--continuum plus the $\rho$ pole,
the remainder a summation over 
additional isovector vector meson poles.

The VMD model of ref.~\cite{Mergell:1996bf} focused more on
fitting the space-like nucleon form factors.
It was found sufficient to use 3 additional vector meson poles,
two of which are chosen to be the physical particles 
$\rho(1450)$ and $\rho(1700)$ and denoted as
$\rho^{\prime\prime}$ and $\rho^{\prime\prime\prime}$.
In Ref.~\cite{Hammer:1996kx}, which extends the scenario 
to include time-like data, the third pole
is left adjustable to compensate for the neglect of higher mass
continuum like $N\bar{N}$. 
The pole mass was determined to be $M_{\rho^\prime}=1.4035$~GeV,
in association with fixing $\Lambda^2=12$~GeV$^2$ and
$Q_1^2=0.35$~GeV$^2$ in Eq. (\ref{deflt}).  
The parameters in Eq. (\ref{lt}) remain unaffected in both fits.
Besides the fictitious $\rho^\prime$, one special feature of the model
is to utilize the freedom in insufficient knowledge of
$\rho^{\prime\prime}$ and $\rho^{\prime\prime\prime}$ 
widths and couplings, which are taken as fit parameters. 
Thus only the $\rho(770)$ is isolated from the summation so that
$\tilde{F}_i^\rho(t)$ contains no parameters that need to be determined.

With $a_i^v$ taken as parameters, Eq. (\ref{asym}),
i.e. PQCD power counting, can be achieved by {\it choosing} 
the residues $a_i^{v}$ of the vector meson poles such that 
the leading coefficients in the $1/t$ expansion cancel. 
This is what the use of a single $\rho$-pole can never achieve, 
since further excited states are needed for such cancellation. 
This also means that one has only an effective {\it model}
since the dynamical parameters for $\rho(1450)$ and $\rho(1700)$
would likely not correspond to physical ones.
We summarize in Table~\ref{MassPoles} the relevant meson poles and the
corresponding residues of the higher excited states given by Table~1 of
Ref.~\cite{Hammer:1996kx} (only the so-called ``Fit 2" is needed).

\begin{table}[t!]
\begin{center}
\caption{
The relevant poles~($M_{v}$ in GeV) and the corresponding residues 
that enter the summation in Eq. (\ref{ffivm}). 
}
\begin{tabular}{cc|cc|cc}
\multicolumn{2}{c|}{$M_{\rho^\prime}=1.4035$}&\multicolumn{2}{c|}{$
M_{\rho^{\prime\prime}}=1.45$}&\multicolumn{2}{c}{
$M_{\rho^{\prime\prime\prime}}=1.69$}\\
\hline
$a_1^{\rho^\prime}$&$a_2^{\rho^\prime}$&
$a_1^{\rho^{\prime\prime}}$&$a_2^{\rho^{\prime\prime}}$&
$a_1^{\rho^{\prime\prime\prime}}$&$a_2^{\rho^{\prime\prime\prime}}$\\
\hline
 $-9.913$ & $-4.731$ & $13.01$ & $0.263$ & $-3.497$ & $2.947$
\label{MassPoles}
\end{tabular}
\end{center}
\end{table}

The parametrization agrees with experiment quite
well up to $t\sim 6$~GeV$^2$, beyond which 
it overruns data because of the choice of low $\Lambda$ in
the logarithmic factor $L(t)$. 
This is in contrast with the empirical fit of Eq.~(\ref{empirical}).
It arises in part because the VMD model focuses more on the spacelike
region where one has more data, but is of less concern to us.
In order to conform to experiment for 
larger timelike momentum transfer, however, 
a modification of the $L(t)$ factor is needed. 
The empirical fit of Eq.~(\ref{empirical}) suggests a 
convenient modification
\[
L\left(t\right)=\left\{
\begin{array}{l}
\biggl[ \ln \biggl( \dfrac{\Lambda^2 - t}{Q_1^2}
\biggr)\biggr]^{-\gamma} 
\ \hfill \mbox{for}\quad t \leq \frac{\Lambda^2}{2}-\Delta,
 \\
-\frac{1}{2\Delta}L^\prime\left(\frac{\Lambda^2}{2}-\Delta\right)
\left(t-\frac{\Lambda^2}{2}\right)^2+H\left(\Delta\right)
\quad\quad\quad  \\
\ \hfill  
 \mbox{for}\quad \vert t- \frac{\Lambda^2}{2}\vert < \Delta,
 \\
\biggl[ \ln \biggl( \dfrac{t}{Q_1^2}
\biggr)\biggr]^{-\gamma} \hfill 
\ \hfill \mbox{for}\,\quad t \geq \frac{\Lambda^2}{2}+\Delta.
\end{array}\right.
\]
where we match a parabola between the interval
$\frac{\Lambda^2}{2}-\Delta<t<\frac{\Lambda^2}{2}+\Delta$ 
by tuning the constant $H(\Delta)$.
Note that $-L^\prime\left(\Lambda^2/2-\Delta\right)=
L^\prime\left(\Lambda^2/2+\Delta\right)$
and $L\left(\Lambda^2/2-\Delta\right)=L\left(\Lambda^2/2+\Delta\right)$,
or $L(t)$ is symmetric with respect to $t=\Lambda^2/2$.
To smooth out the artificial rise that occurs beyond 
$t\sim 6$~GeV$^2=\Lambda^2/2$ for the original fit, 
we choose $\Delta=0.5$ GeV$^2$ and $H\left(\Delta\right)\sim0.10279.$

\begin{figure}[t!]
\begin{center}
\epsfig{figure=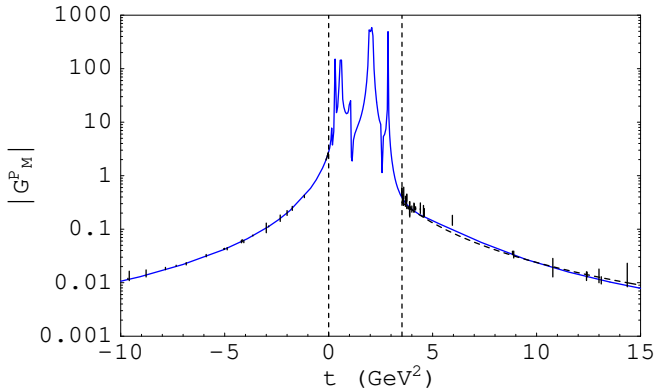,width=3.4in}
\caption{
The VMD-based proton magnetic form factor with the modified
logarithmic factor. 
The unphysical region is marked by two vertical lines.
The dash-line in the time-like region
represents
a fit with $C=53$ in Eq.~(\ref{empirical}).} \label{modified}
\end{center}
\end{figure}

Fig.~\ref{modified} shows the resulting
VMD-based proton magnetic form factor 
for both space-like and time-like $t$,
with the modified log factor in the time-like region,
where the result is plotted in more detail in Fig.~\ref{proton}.
We see that the trend of the proton data can be described by this model. 
In contrast, from Fig.~\ref{neutron},
where the extension of the VMD result to neutron case is given, 
there is a significant deviation from $G_M^n$ data, 
signaling the incapability of the three-plus-one pole fit 
to describe the neutron data consistently.
This fact was admitted to in Ref.~\cite{Hammer:1996kx}.

One useful aspect of a dispersion approach is that 
analyticity can help determine 
the signs of the time-like form factors from space-like region.
Unlike Eqs.~(\ref{fitP}) and (\ref{fitN}) where 
the signs are put in by hand,
it is more natural in the dispersion analysis 
since all the time-like information can in principle be 
obtained via the dispersion relation in Eq.~(\ref{disp}),
where the spectral function Im$\,F(t)$ is
analytically continued from the space-like region.
One can readily check this by finding out the value of the
magnetic form factors from the VMD analysis at threshold:
$G^p_M(4m_N^2)\cong +0.39$ while $G^n_M(4m_N^2)\cong -0.22$.
From both Figs. \ref{proton} and \ref{neutron}, 
since neither $G^p_M$ nor $G^n_M$ seem to cross the $t$-axis,
$G^p_M$ remains positive while $G^n_M$
remains negative throughout the timelike region.
VMD analysis thus gives $G^p_M$ and $G^n_M$ with a relative sign.

\subsection{$B^0\to D^{*-}p\bar n$ in VMD Approach}

We calculate the branching fraction of $B^0\rightarrow D^{*-}p\bar{n}$ 
through the vector portion of the charged weak current in the VMD model,
again taking $a_1$ as extracted from $B^0\rightarrow D^{*-}\rho^+$.
The results for $B^0\to D^{*-}p\bar n$ are,
\begin{eqnarray}
{\rm Br}_V^{{\rm LF}}=(1.69\times
10^{-4})\left(\frac{a_1}{0.74}\right)^2\,,\\
{\rm
Br}_V^{{\rm BSW}}=(2.06\times10^{-4})\left(\frac{a_1}{0.86}\right)^2\,,
\end{eqnarray}
%
which are respectively about $12\%$ and $14\%$ of 
the experimental value of $(14.5{+3.4\atop -3.0}\pm2.7)\times 10^{-4}$.
Varying pole masses slightly does not drastically modify the result.
The differential decay rate is plotted in Fig.~\ref{dGdq},
where one sees that it peaks not far above 
the $p\bar{n}$ threshold of $t\sim 3.53$~GeV$^2$. 
The contribution from the region of $3.53<t<6$~GeV$^2$ is 
more than $80\%$ of the total rate for both LF and BSW models. 
Had we used Eq. (\ref{deflt}),
because of the artificial rise in this original logarithmic factor, 
the contribution from $t>6$~GeV$^2$ ($m_{p\bar n} > 2.45$ GeV) 
would be $\sim$ 2.5 times higher.
The contribution from $t<6$~GeV$^2$, however, is unchanged.

It is clear that the branching fractions obtained in 
the VMD approach of Ref.~\cite{Hammer:1996kx}
are typically 5 times smaller than our phenomenological model 
discussed in the previous section.
This can be simply traced to the inadequacy in
accounting for neutron data by the VMD model,
as seen by comparing Figs.~\ref{proton} and~\ref{neutron}. 
While giving a reasonable fit in the proton case, 
the absolute value of its neutron form factor 
is simply below all data points.
The VMD model was tuned more on the proton 
where data is much more abundant.
Since the proton and neutron magnetic form factors are opposite in sign,
if the VMD approach is improved to give better account of $|G^n_M|$ data, 
the combination $G^P_M-G^n_M$ would increase.

\begin{figure}[t!]
\begin{center}
\epsfig{figure=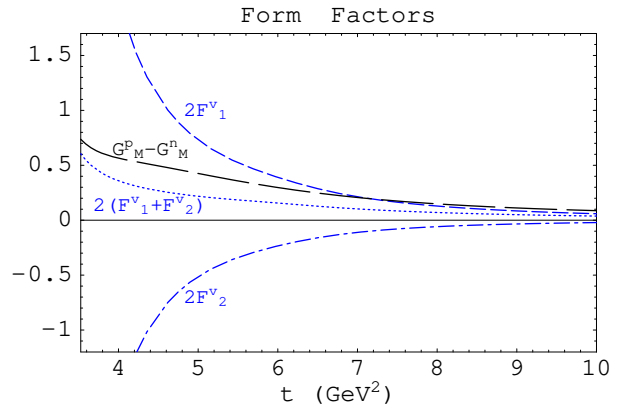,width=3.4in}
\end{center}
\caption{
$G^p_M-G^n_M$~(long-dashed) from our fit to nucleon form factor data 
compared with $2F^{v}_1+2F^{v}_2 (=G^p_M-G^n_M$,~dotted) 
from the VMD analysis, 
for the kinematically allowed region of $t$ in $B^0\to D^{*-} p \bar n$.
Also shown are $2F^{v}_1$~(short-dashed) and
$2F^{v}_2$~(dot-dashed) from the VMD model.
} \label{FormFactors}
\end{figure}

\section{Discussion and Applications}\label{discussions}

\subsection{Nucleon Form Factor Decompositions}

We plot various form factor combinations in Fig.~\ref{FormFactors}.
The long dashed line shows 
$G^p_M-G^n_M=2(F^v_1+F^v_2)$ from our phenomenological fit,
which assumes $G^p_M$ and $G^n_M$ have opposite sign.
As discussed in the previous section,
because the VMD model gives much lower value for $\vert G^n_M\vert$,
$G^p_M-G^n_M$ in the VMD model 
(denoted as dotted line and labeled by $2(F^v_1+F^v_2)$)
stays below the phenomenological model.
However, had we chosen the proton and neutron form factors 
to be of the same sign , $G^p_M-G^n_M$ for our phenomenological fit
would be very close to the solid line which is the $t$-axis, 
and would give a rate that is two orders of magnitude too small.

\begin{table}[t!]
\caption{
Comparison of nucleon form factor decompositions.
The fraction ${\cal R}_X\equiv{\cal B}_X/$Br$_V$ are defined in text, 
where $X$ stands for any combination of $F_1$, $F_2$ or their product.
}
\begin{center}
\begin{tabular}{cccc}
& ${\cal R}_{(F^v_1+F^v_2)}$ & ${\cal
R}_{F^v_2}$ & ${\cal R}_{(F^v_1+F^v_2)\cdot
F^v_2}$ \\
\hline
LF/BSW & 1.3 & 0.24 & $-0.54$ \\
\hline\hline
& ${\cal R}_{F^v_1}$ & ${\cal R}_{F_2^{v}}$ &
${\cal R}_{F^v_1\cdot F^v_2}$ \\
\hline
LF/BSW & 26.00 & 18.25 & $-43.25$  \\
\end{tabular}
\end{center}
\label{decomposition}
\end{table}

Besides helping to fix the sign of $G^n_M$ by analyticity,
another utility of discussing the VMD approach is that 
it can give some insight to the $F_2$ nucleon form factor.
Because of lack of experimental information,
we have dropped the $F_2$ contribution in our phenomenological approach,
and we need to check the validity of this.
The weak vector current induced decay amplitude, 
upon squaring, can be expressed as
\begin{eqnarray}
\left|{\cal M}_V\right|^2&=&\left|{\cal
M}_{F_1+F_2}\right|^2+\left|{\cal M}_{F_2}\right|^2\nonumber\\
& &\quad+2{\rm Re}\left({\cal
M}_{F_1+F_2}{\cal M}^*_{F_2}\right)
\label{anatomy}
\end{eqnarray}
where ${\cal M}_{F_1+F_2,F_2}$ denote $F^v_1+F^v_2,\,F^v_2$
terms in Eq.~(\ref{amplitude}). 
Decomposing
Br$_V={\cal B}_{F_1+F_2}+{\cal B}_{F_2}+{\cal B}_{(F_1+F_2)F_2}$
where the last term is the interference term,
we define the relative fractions such as 
${\cal R}_{(F_1+F_2)}\equiv{\cal B}_{(F_1+F_2)}/$Br$_V$
from $(F^v_1+F^v_2)$ alone.  
We find 
${\cal R}_{(F_1+F_2)},{\cal R}_{F_2},{\cal R}_{(F_1+F_2)\cdot F_2}
 = 130\%,\,24\%,\,-54\%$ (Table~\ref{decomposition}) in the VMD model,
for both LF and BSW $B^0 \to D^*$ form factor models.
Note that the $2(F^v_1+F^v_2) = G^p_M-G^n_M$ term
gives the dominant contribution, which supports the 
approximation used in Sec. II. 
The $F^v_2$ contribution is much smaller even though 
$|F^v_2(t)|$ is greater than $|F^v_1+F^v_2|$ for $t<5$ GeV$^2$,
as shown by the dot-dash curve in Fig. \ref{FormFactors}. 
The interference term contributes $\sim 40\%$ 
of the $F^v_1+F^v_2$ contribution. 
The destructive nature is due to the relative sign between 
$F^v_1+F^v_2$ and $F^v_2$, which reduces the combined effect of
the last two terms in Eq.~(\ref{anatomy}).

It is instructive to compare with Eq.~(\ref{weak_current}),
where one decomposes into $F_1$ and $F_2$ directly.
As shown in Table~\ref{decomposition}, 
the individual terms from this decomposition
are an order of magnitude larger, 
and only after strong cancellations does one arrive at Br$_V$.
It is therefore not a very useful decomposition.
We see from Fig.~\ref{FormFactors} that 
the magnitudes of $F^v_1$ and $F^v_2$ are all greater than their sum, 
hence $F^v_1 + F^v_2$ together with $F^v_2$ is a better decomposition
for $t$ close to threshold,
which is what we used in our phenomenological study.

\begin{table}[t!]
\caption{
Branching fractions of 
the $B\to \bar D^{(*)}p\bar{n}$ modes from the vector current, 
obtained by using the phenomenological form factors with
$|G^p_M|=|G^p_E|$ and $|G^n_M|=|G^n_E|$ ($|G^n_E|=0$)
for the first (second) set.
}
\begin{center}
\begin{tabular}{lcccc}
                  &\multicolumn{2}{c}{Br$_V\times 10^4$}
                  &\multicolumn{2}{c}{Br$_V\times 10^4$} \\
\raisebox{1.5ex}{} & \multicolumn{2}{c}{$|G^n_M|=|G^n_E|$} &
                     \multicolumn{2}{c}{$|G^n_E|=0$} \\
\raisebox{1.5ex}{} & LF & BSW & LF & BSW \\
\hline
$B^0\rightarrow D^{*-}p\bar{n}$ & $7.14^{+0.69}_{-0.65}$ &
$8.72^{+0.85}_{-0.79}$ & $8.96^{+1.02}_{-0.94}$ & $10.94^{+1.25}_{-1.15}$
\\
$B^+\rightarrow \bar{D}^{*0}p\bar{n}$ & $7.64^{+0.74}_{-0.69}$ &
$9.33^{+0.90}_{-0.85}$ & $9.59^{+1.10}_{-1.01}$ & $11.71^{+1.33}_{-1.23}$
\\
\hline
$B^+\rightarrow \bar{D}^0 p\bar{n}$ & $3.92^{+0.39}_{-0.35}$ &
$3.21^{+0.32}_{-0.28}$ & $4.89^{+0.58}_{-0.51}$ & $4.00^{+0.47}_{-0.42}$
\\
$B^0\rightarrow D^-p\bar{n}$  & $3.66^{+0.36}_{-0.32}$ &
$2.99^{+0.30}_{-0.26}$ & $4.56^{+0.54}_{-0.48}$ & $3.73^{+0.44}_{-0.39}$
\\
\end{tabular}
\label{BtoDpn}
\end{center}
\end{table}

\subsection{Prediction for $B\to D^{(*)}p\bar{n}$ Modes}

\begin{figure}[b!]
\begin{center}
\epsfig{figure=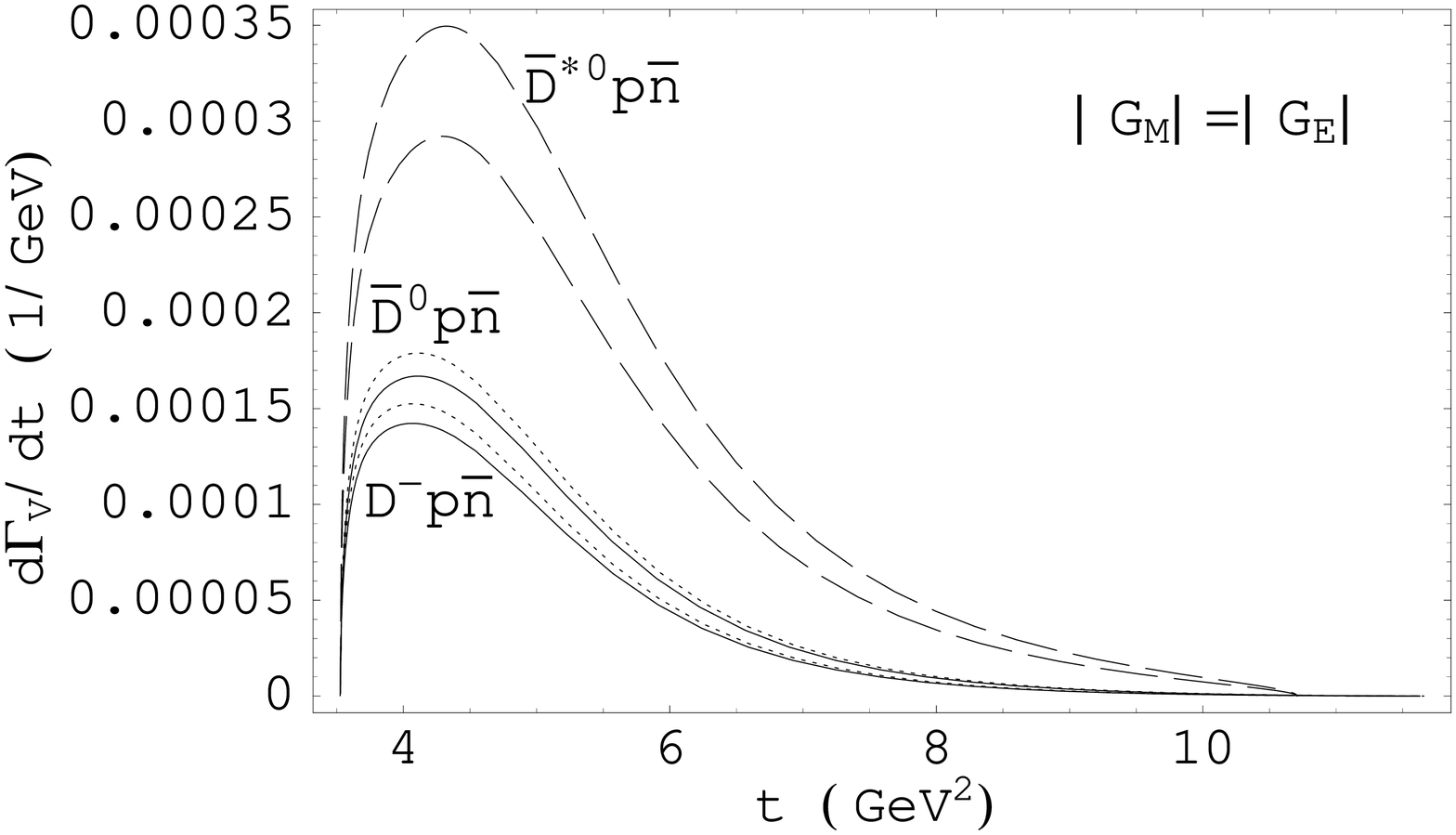,width=3.4in}
\end{center} 
\caption{
The differential decay rates arising from nucleon vector current.
Upper (lower) dashed line is for $B^0 \rightarrow \bar{D}^{*0}p\bar{n}$ 
with the BSW (LF) hadronic form factors; 
upper (lower) dotted and solid lines
are for the $B^0 \rightarrow \bar{D}^0 p\bar{n}$
and $B^0\rightarrow D^- p\bar{n}$ modes using the LF (BSW) model.
}
\label{dGdqV_others}
\end{figure}

Our phenomenological approach can be applied to the modes of
$B^+\to \bar{D}^{(*)0}p\bar{n}$ and $B^0\to D^-p\bar{n}$. 
We show in Table~\ref{BtoDpn} the results based on the vector current 
with the $B^0\to D^{*-}p\bar{n}$ mode included for comparison.
The differential decay rates for 
the other three $B\to D^{(*)}p\bar{n}$ decays
are given in Fig.~\ref{dGdqV_others}.
We have used the central values of the effective coefficients 
that are extracted from the two-body modes~\cite{Cheng:1999kd} 
$\bar{B}^0\to D^{*+}\rho^-$, $D^+\rho^-$ with values 
$a_1^{LF(BSW)}=0.74\pm 0.18\pm0.06~(0.86\pm0.21\pm0.07)$ and
$a_1^{LF(BSW)}=0.89\pm0.08\pm0.07~(0.91\pm0.08\pm0.07)$ respectively.
We note that the $\bar D^*$ modes in Table~\ref{BtoDpn} 
are close in rate, and likewise for $\bar D$ modes,
which is easy to understand because of simple replacement of 
spectator quark in $B\to \bar D^{(*)}$ transition.

Although only the contribution from the weak vector current 
can so far be calculated, we can estimate the values of these 
other baryonic modes by following the recipe of Sec. II.
Inspection of Table~\ref{BtoDpn} shows that the ratio of 
${\rm Br}_V (B^+\rightarrow \bar{D}^{*0}p\bar{n})/
{\rm Br}_V(B^0\rightarrow \bar{D}^{*-}p\bar{n})$ remains fixed
regardless of $B\rightarrow \bar{D}^{*}$ form factor models.
Assuming similar behavior for axial contribution,
we expect the $B^+\rightarrow \bar{D}^{*0}p\bar{n}$
branching fraction to be scaled by the same factor 
and find the value of
$\sim 15.5\times10^{-4}$ that is only slightly larger than 
the $B^0\to D^{*-}p\bar{n}$ mode, as given in Table~\ref{BrExp}.
The predicted branching fractions for 
$B^+\to \bar{D}^0p\bar{n}$ and $B^0\to D^-p\bar{n}$
from the same ansatz are in general are 2--3 times smaller.
Since one is comparing $B\to \bar D$ vs. $B\to \bar D^*$,
there is stronger model dependence on the transition form factor.

\begin{table}[t!]
\caption{
Branching fractions estimated by scaling 
${{\rm Br}_V (B\rightarrow D^{(*)}p\bar{n})}
/{{\rm Br}_V(B^0\rightarrow D^{*-}p\bar{n})}$ 
of Table~\ref{BtoDpn} with respect to 
the ${\rm Br}(B^0\rightarrow D^{*-}p\bar{n})$ 
experimental central value of $14.5\times 10^{-4}$ (Eq. (1)).
}
\begin{center} \begin{tabular}{lcccc}
                  &\multicolumn{2}{c}{Br$\times 10^4$}
                  &\multicolumn{2}{c}{Br$\times 10^4$} \\
\raisebox{1.5ex}{} & \multicolumn{2}{c}{$|G^n_M|=|G^n_E|$} &
                     \multicolumn{2}{c}{$|G^n_E|=0$} \\
\raisebox{1.5ex}{} & LF & BSW & LF & BSW \\
\hline
$B^+\rightarrow \bar{D}^{*0}p\bar{n}$ & $15.52$ & $15.51$ & $15.52$ & 
$15.51$
\\
$B^+\rightarrow \bar{D}^0 p\bar{n}$ & $7.97$ & $5.34$ & $7.92$ &
$5.30$
\\
$B^0\rightarrow D^-p\bar{n}$  & $7.43$ & $4.98$ & $7.38$ & $4.94$
\\
\end{tabular}
\label{BrExp}
\end{center}
\end{table}

One can see from both Tables~\ref{BtoDpn} and \ref{BrExp} 
as well as Fig.~\ref{dGdqV_others} that the LF results are 
generally larger than the BSW ones for $B\to \bar{D}$ modes, 
but the case is reversed for the $B\to \bar D^{*}$ modes. 
This can be understood from the differences in $a_1$ and
hadronic form factors of the LF and BSW cases.
For the $B\to \bar{D}$ modes, the only hadronic form factor involved 
is $F_1^{BD}$, which behaves as a dipole and monopole in 
the LF and BSW models, respectively. 
This leads to $F_1^{LF}(t)>F_1^{BSW}(t)$ in the physical timelike region
while $a_1^{LF} \cong a_1^{BSW} \cong 0.9$, giving a larger Br$_V^{LF}$.  
In contrast, for $B\to \bar D^{*}$ modes, 
although the dominant hadronic form factor 
$A_1^{LF}>A_1^{BSW}$ in the physical timelike region, 
it behaves as monopole for both LF and BSW models,
hence the difference between $A_1^{LF}$ and $A_1^{BSW}$ is not 
as large as the previous case.
However, $a_1^{LF} \sim 0.74 < a_1^{BSW} \sim 0.86$
hence we have the opposite result that BSW rates are bigger.

\subsection{Predictions for $B^0\to D^{*-}+$ Strange Baryons}

Our phenomenological approach can be extended to 
$B^0$ decay into $D^{*-}$ plus baryon pairs containing strangeness.
Recall that in Sec.~\ref{fact} we utilized SU(2) symmetry to obtain 
the relation $F^W_{1,2}=F^p_{1,2}-F^n_{1,2}$ for the $p\bar n$ mode. 
In the SU(3) limit we can~\cite{Georgi:1984kw} extend this relation 
to modes containing strange baryons.
Starting from Eq.~(\ref{weak_current}),
denoting $F^W_{1,2}({\rm \bf B}_s\bar{\rm \bf B}_s^\prime)$
as the weak form factor that appears in the matrix element
$\langle {\rm \bf B}_s\bar{\rm \bf B}_s^\prime|V^+_{\mu}|0\rangle$,
we find
\begin{eqnarray}
F^W_{1,2}(\Sigma^{+,0}\bar\Sigma^{0,-})&=&
\mp \sqrt{2}\left(F^p_{1,2}(t)+\frac{1}{2}\,F^n_{1,2}(t)\right),
\nonumber\\
F^W_{1,2}(\Xi^0\bar\Xi^-)&=&
F^p_{1,2}(t)+2\,F^n_{1,2}(t),
\nonumber\\
F^W_{1,2}(\Sigma^+\bar\Lambda)&=&
-\sqrt{\frac{3}{2}}\,F^n_{1,2}(t).
\end{eqnarray}
These relations enable us to calculate the contributions from 
the vector currents to the strange baryonic modes.
We shall only give results from the phenomenological approach 
with $|G^n_E|=|G^n_M|$.

\begin{figure}[t!]
\begin{center}
\epsfig{figure=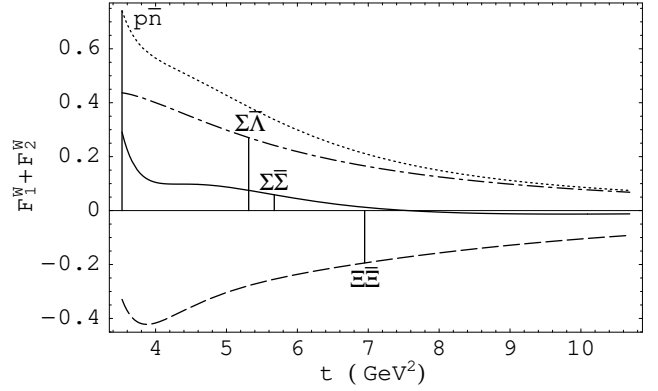,width=3.3in}
\end{center}
\caption{
$F^W_1+F^W_2$ 
for the $p\bar n$~(dotted), $\Sigma^0\bar\Sigma^-$~(solid), 
$\Xi^0\bar\Xi^-$~(dashed) and $\Sigma^+\bar\Lambda$~(dot-dashed) modes,
respectively. 
The vertical lines specify the thresholds for the baryonic pairs.
}
\label{ff}
\end{figure}

\begin{figure}[b!]
\begin{center}
\epsfig{figure=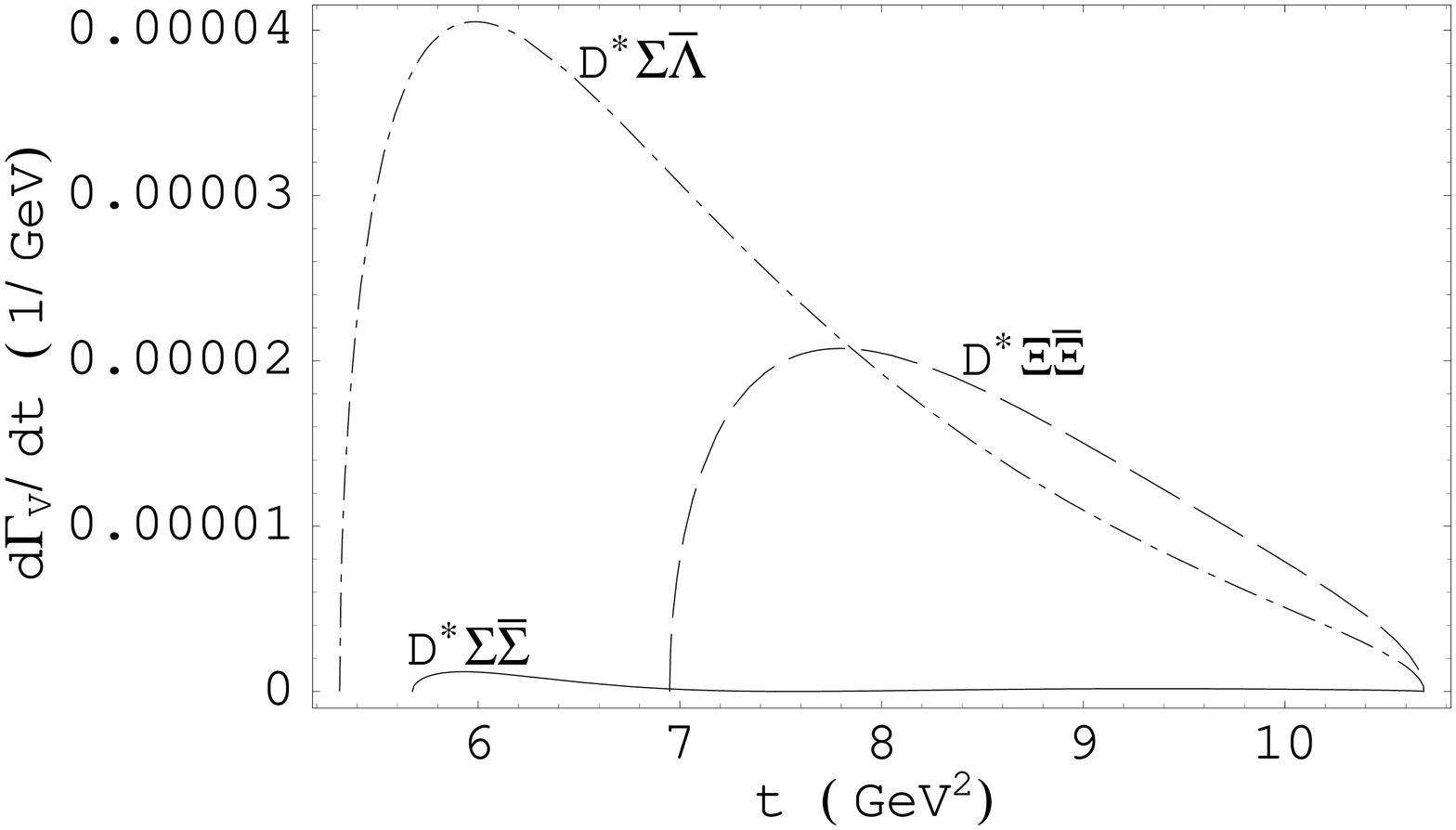,width=3.3in}
\end{center}
\caption{
Differential rates of $B^0\rightarrow
D^{*-}\Sigma^+\bar{\Lambda}$~(dot-dashed), $D^{*-} \Xi^0
\bar{\Xi}^-$~(dashed) and $D^{*-}\Sigma^+\bar{\Sigma}^0$~(solid) 
from the phenomenological model, with LF hadronic form factors.
}
\label{SU3_LF}
\end{figure}

The strange baryon modes all have rates smaller than
that of the $p\bar{n}$ mode due to the following reason.
In Fig.~\ref{ff} we plot the form factor combination
$F^W_1+F^W_2=2(F^v_1+F^v_2)$ of Eq.~(\ref{amplitude}).
One can see that the largest value is at the $p\bar n$ threshold, 
while for the strange baryonic modes 
the corresponding threshold values are all smaller.
Reading off from Fig.~\ref{ff}, it is clear that
the $\Sigma^+\bar\Lambda$ mode would be 
the dominant strange baryonic mode, 
with the two $\Sigma\bar\Sigma$ modes the smallest.
This is shown in Fig.~\ref{SU3_LF} for 
the differential decay rates using LF mesonic form factors,
and the total decay rates given in Table~\ref{Br}.
Note that the differential decay rate for the $D^{*-}\Sigma\bar\Sigma$
mode has a zero at $t\sim 7.5$~GeV$^2$ because 
$F^W_1(\Sigma\bar\Sigma)+F^W_2(\Sigma\bar\Sigma)$ changes sign, 
as can be seen from Fig.~\ref{ff}.

\begin{table}[t!]
\begin{center}
\caption{
Branching fractions for $B^0\to D^{*-}+$ strange baryon pairs 
from vector current using the phenomenological 
$G_M$ form factors of Sec.~\ref{fact} assuming $|G^n_E|=|G^n_M|$.
}
\begin{tabular}{lcc}
                  & Br$_V^{{\rm LF}}\times 10^4$
                  & Br$_V^{{\rm BSW}}\times 10^4$ \\
                  \cline{2-3}
\hline
$B^0\rightarrow D^{*-}p\bar{n}$ & 7.19 & 8.78 \\
$B^0\rightarrow D^{*-} \Sigma^+\bar\Lambda$ & 1.09 & 1.38 \\
$B^0\rightarrow D^{*-}\Xi^0\bar{\Xi}^-$& 0.52 & 0.66 \\
$B^0\rightarrow D^{*-} \Sigma^+\bar{\Sigma}^0 $ & 0.01 & 0.02 \\
$B^0\rightarrow D^{*-} \Sigma^0\bar{\Sigma}^-$ & 0.01 & 0.02 \\
\end{tabular}
\label{Br}
\end{center}
\end{table}

\section{Conclusion}\label{Conclusions}

The main result of this paper is an attempt to account for
CLEO's result on $B^0\rightarrow D^{*-}p\bar{n}$.
In a factorization approach, the nucleon pair is viewed as 
produced directly from the weak current.
We then use an isospin relation of weak and em vector currents
to obtain the vector weak nucleon form factors directly from their em
partners, where a relatively large database exists.
It is interesting that we can account for 
$\sim 60\%$ of the observed Br$(\btodpn)$ rate in this way.
A VMD model that attempts at fitting nucleon form factor data
was discussed to clarify certain issues.

Interference of the weak vector current induced amplitude
with the amplitude induced by the weak axial current 
does not manifest itself in the total rate. 
The latter is a simple sum of the absolute squares of both
vector and axial vector. 
A rough estimate of axial vector contribution,
together with the dominant vector contribution, 
seem to fit the measured rate.
However, for a more reliable calculation,
more measurements on time-like region $G^{p,n}_E$ and $g_A$ are urged.

We apply our analysis to $B^+\to \bar{D}^{(*)0}p\bar{n}$
and $B^0\to D^-p\bar{n}$ modes to get the rates
arising from the vector current. 
We find
Br$(B^+\to \bar{D}^{*0} p\bar{n})\sim$ Br$(B^0\to D^{*-} p\bar{n})$ 
and Br$(B^+\to \bar{D}^0p\bar{n})\sim$ Br$(B^0\to D^-p\bar{n})$,
with the latter modes having smaller
rates slightly below the $10^{-3}$ level.
Our analysis is also applied to baryonic modes that contain strangeness.
The estimated branching fractions are generally
lower than the $p\bar{n}$ mode due to smaller couplings and
higher thresholds.
The largest modes, $B^0\rightarrow D^{*-}\Sigma^+\bar{\Lambda}$, 
is predicted at the $1\times 10^{-4}$ level.

For the analogous picture of $B\to\eta^\prime \bar\Lambda p,\ 
\gamma\bar\Lambda p$ as descending from $B\to \eta^\prime K$ 
and $K^*\gamma$ via $K^{(*)}$ exchange, 
the baryon pairs are not produced by the $W$ boson,
and it seems that the approach used here cannot be readily applied.
However, some experience obtained may still be valuable.
For example, in the VMD approach to $B\to \bar D^* p\bar n$,
more than one pole and cancellations among them are required 
to reproduce the QCD predicted asymptotic behavior.
For the charmless cases, the baryon pairs are 
no longer produced by the charged current.
Instead, the $K^{(*)}$ resonances that correspond to
$\rho^{\prime,\prime\prime,\prime\prime\prime}$ in VMD approach,
appear more as string excitations.
They need not obey the same QCD power countings,
and perhaps may result in larger rates.

Finally, it is of great interest to estimate 
the rate of the charmless baryonic mode $B^0\to \rho^- p\bar{n}$ 
by replacing $D^{*-}$ in the Feynman diagram depicted in 
Fig.~\ref{Ansatz}(b) with $\rho^-$.
Since this is a tree-dominant mode, 
extending from the study presented in this paper,
Br$(B^0\to \rho^- p\bar{n})$ could be as large as that of the
two-body mode $B^0\to \rho^-\rho^+$, 
analogous to the relative strength of
Br$(B^0\to D^{*-} p\bar{n})$ vs Br$(B^0\to D^{*-}\rho^+)$. 
Estimating via Br$(B^0\to \rho^- p\bar{n})\sim$ Br$(B^0\to \rho^-\rho^+)
\times$ Br$(B^0\to D^{*-} p\bar{n})/$Br$(B^0\to D^{*-}\rho^+)$.
From Br$(B^0\to \rho^-\rho^+)\sim $ (20--40) $\times
10^{-6}$~\cite{Ali:1998eb,Hou:2000tf} we get Br$(B^0\to \rho^-
p\bar{n})\sim$ (0.4--0.8) $\times 10^{-5}$.
Alternatively,
we can scale from $B\to \bar D^* p\bar n$ by $\vert V_{ub}/V_{cd}\vert^2$
and phase space, decay constant etc., and again find 
$B\to \rho^- p\bar n$ at $10^{-5}$ order.
Charmless decays are of great current interest.
A more detailed discussion of $B\to \rho p\bar n$ modes
is given elsewhere \cite{Chua:2001xn}
.

\acknowledgments

We thank S.\ J.\ Brodsky for illuminating discussions
and useful suggestions. 
This work is supported in part by 
the National Science Council of R.O.C. under Grants
NSC-89-2112-M-002-063, NSC-89-2811-M-002-0086 and NSC-89-2112-M-002-062,
the MOE CosPA Project, and the BCP Topical Program of NCTS.


%

\end{document}